\documentclass[manuscript]{aastex631}
\usepackage{amsmath,amstext}
\usepackage{graphicx}
\usepackage{natbib}
\usepackage{amssymb}
\usepackage{makecell}
\usepackage{multirow}
\usepackage{threeparttable}
\def\vec#1{\ensuremath{\mathbf{#1}}}
\defcitealias{2021ApJ...921L..17G}{G21}

\received{}
\revised{}
\accepted{}

\shortauthors{Shi et al.}
\begin{document}

\title{Influence of fine structures on gyrosynchrotron emission of flare loops modulated by sausage modes}

\correspondingauthor{Mijie Shi}
\email{shimijie@sdu.edu.cn}

\author{Mijie Shi}
\affiliation{Shandong Key Laboratory of Optical Astronomy and Solar-Terrestrial Environment, School of Space Science and Physics, Institute of Space Sciences, Shandong University, Weihai, Shandong, 264209, China}

%\author{Tom Van Doorsselaere}
%\affiliation{Centre for mathematical Plasma Astrophysics, Department of Mathematics, KU Leuven, B-3001 Leuven, Belgium}

\author{Bo Li}
\affiliation{Shandong Key Laboratory of Optical Astronomy and Solar-Terrestrial Environment, School of Space Science and Physics, Institute of Space Sciences, Shandong University, Weihai, Shandong, 264209, China}

\author{Mingzhe Guo}
\affiliation{Shandong Key Laboratory of Optical Astronomy and Solar-Terrestrial Environment, School of Space Science and Physics, Institute of Space Sciences, Shandong University, Weihai, Shandong, 264209, China}



\begin{abstract}
Sausage modes are one leading mechanism for interpreting short period 
	quasi-periodic pulsations (QPPs) of solar flares.
Forward modeling their radio emission
	is crucial for identifying sausage modes observationally and 
	for understanding their connections with QPPs.
Using the numerical output from three-dimensional magnetohydrodynamic (MHD) simulations, 
	we forward model the gyrosynchrotron (GS) emission of flare loops modulated by sausage modes
	and examine the influence of loop fine structures.
The temporal evolution of the emission intensity is analyzed 
	for an oblique line of sight crossing the loop center.
We find that the low- and high-frequency intensities oscillate in-phase at the period of sausage modes
	for models with or without fine structures.
For low-frequency emissions where the optically thick regime arises,
	the modulation magnitude of the intensity is dramatically reduced by the fine structures at some viewing angles.
On the contrary, for high-frequency emissions where the optically thin regime holds,
	the effect of fine structures or viewing angle is marginal.
Our results show that the periodic intensity variations of sausage modes
	are not wiped out by the fine structures,
	and sausage modes remains a promising candidate mechanism for QPPs even when flare loops are fine-structured.

\end{abstract}




\section{Introduction} 
\label{S-Introduction}
Quasi-periodic pulsations (QPPs) refer broadly to the oscillatory intensity variations commonly observed 
	in solar flare emissions across a broad range of passbands
	\citep[see the reviews by, e.g.,][]{2009SSRv..149..119N,2020STP.....6a...3K}.
In spite of the abundant observed instances, 
	the physical mechanisms responsible for QPPs still remain inconclusive
	\citep{2016SoPh..291.3143V,2021SSRv..217...66Z}.
Sausage modes can cause periodic compression and rarefaction of flare loops and 
	are thus thought of as one of the mechanisms accounting for QPPs in solar flares
	\citep[see the recent review by][]{2020SSRv..216..136L}.
In terms of observations, candidate sausage modes have been reported
	in the radio band \citep[e.g.,][]{2005A&A...439..727M,2015A&A...574A..53K},
	in the extreme-ultraviolet (EUV) \citep[e.g.,][]{2012ApJ...755..113S,2016ApJ...823L..16T}, 
	as well as in X-ray \citep[e.g.,][]{2010SoPh..263..163Z}.

The classical theory of sausage modes \citep{1983SoPh...88..179E} assumes that sausage modes are supported 
	by an axisymmetric monolithic loop.
However, high resolution observations 		
	\citep[e.g.,][]{2012ApJ...755L..33B,2013Natur.493..501C,2013A&A...556A.104P,2017ApJ...840....4A}
	suggest that coronal loops are fine-structured or multi-stranded.
Multi-stranded loop models have been invoked 
	to explain such observations as the coronal fuzziness
	\citep{2009ApJ...694.1256T,2010ApJ...719..576G} and 
	the time lag of EUV light curves
	\citep{2003ApJ...593.1174W,2012ApJ...753...35V}.
Transverse kink oscillations in multi-stranded loops have attracted substantial attention
	in both analytical studies \citep[e.g.,][]{2008ApJ...676..717L,2008A&A...485..849V,2010ApJ...716.1371L} 
	and MHD simulations \citep[e.g.,][]{2008ApJ...679.1611T,2009ApJ...694..502O,2011ApJ...731...73P,2016ApJ...823...82M,2019ApJ...883...20G}.
For flare loops, fine strands tend to be common to see as well
	\citep[e.g.,][]{2013AstL...39..267Z,2016ApJ...823L..16T}.
This led \citet[][hereafter \citetalias{2021ApJ...921L..17G}]{2021ApJ...921L..17G}
	to examine the influence of fine structures on sausage modes in flare loops,
	the primary conclusion being that the global sausage mode is still identifiable in spite of the loop fine structures.

One key question to answer is then whether the fine structures can influence the 
	emissions from multi-stranded flare loops experiencing sausage perturbations.
	This issue is particularly necessary to address for
	the radio band, where high temporal resolution can be achieved.
For solar radio emissions, the gyrosynchrotron (GS) emission is known to dominate at the millimeter and
	centimeter wavelengths \citep{1998ARA&A..36..131B}.
GS emissions of flare loops can be modulated by the sausage modes therein.
In turn, these modulations also provide signatures for identifying the sausage modes in flare loops.
Along this line of thinking, a variety of forward modeling analyses with different levels of sophistication have been performed,
	from the model where a homogeneous emission source was assumed
	\citep[e.g.,][]{2006A&A...446.1151N,2007ARep...51..588R,2008ApJ...684.1433F,2012ApJ...748..140M},
	to the models where inhomogeneity
	\citep[e.g.,][]{2014ApJ...785...86R,2015A&A...575A..47R} or loop curvature \citep[e.g.,][]{2015SoPh..290.1173K} was taken into account.

In this work, we forward model the GS emission of flare loops modulated by sausage modes.
New is that we take into account the fine structures of flare loops and 
	examine the influence of fine structures on the GS emission.
Section \ref{S-setup} shows the numerical model.
In Sections \ref{S-results}  and \ref{further computations} we present the forward modeling results.
Section \ref{S-summary} summarizes this study.

\section{Numerical Model}
\label{S-setup}
The MHD model that we use for forward modeling purposes was
	described by \citetalias{2021ApJ...921L..17G},
    where three-dimensional time-dependent simulations were performed to examine
    how fast sausage modes are influenced by fine structures in
    straight, field-aligned, flare loops.
Two MHD models, to be labeled `noFS' and `FS', were constructed
	in \citetalias{2021ApJ...921L..17G}.
For both models,
	the equilibrium magnetic field $\vec{B}$ is $z$-directed, 
	and all equilibrium quantities are $z$-independent.
For model noFS, the loop is axisymmetric with its density prescribed by
\begin{eqnarray}
\rho_{\rm noFS} = \rho_{\rm e}+(\rho_{\rm i} - \rho_{\rm e})f(x,y),
\end{eqnarray}
	where $[\rho_{\rm i},\rho_{\rm e}]$=$[5\times 10^{10},0.8\times 10^{9}]m_p~{\rm cm}^{-3}$
	represent the mass densities at the loop axis and infinitely far from the loop, respectively.
	In addition, $m_p$ is the proton mass.
A function $f(x,y) ={\rm exp}[-(r/R)^{\alpha}]$
	is used to control the density profile,
	with $r=\sqrt{x^2+y^2}$, $\alpha=5$,
	and the nominal loop radius $R=5$~Mm.
The temperature distribution follows the same functional form as the density,
	with the temperature at the loop axis being $T_i=10$~MK
	and that far from the loop being $T_e=2$~MK.
The magnetic field ($B_z$) is prescribed in such a way that transverse force balance is maintained, 
	and the resulting $B_z$ increases from $50$~G at the axis to $77.3$~G far from the loop.
The length of the flare loop is $L=45$~Mm.

Model FS modifies model noFS by introducing fine structures as randomly distributed, 
	small-scale, density variations to the loop interior,
\begin{eqnarray}
\rho_{\rm FS}(x,y) = \rho_{\rm noFS}(x,y) + (\rho_{\rm i}-\rho_{\rm e})f(x,y)g(x,y),
\end{eqnarray}
where
\begin{eqnarray}
g(x,y) = \frac{\sum_{j=1}^{N_{\rm FS}}[{\rm exp}(-\bar{r}_j^{\alpha}){\rm cos}(\pi \bar{r}_j)]}{|\sum_{j=1}^{N_{\rm FS}}[{\rm exp}(-\bar{r}_j^{\alpha}){\rm cos}(\pi \bar{r}_j)]|_{\rm max}},
\end{eqnarray}
with
\begin{eqnarray}
\bar{r}_j = \frac{\sqrt{(x-x_j)^2+(y-y_j)^2}}{R_{\rm FS}}.
\end{eqnarray}
In the above equations,
	$R_{\rm FS}=0.8$~Mm is the nominal radius of fine strands,
	and $[x_j,y_j]$ represents the randomly generated position for an individual strand.
We take the number of fine structure to be $N_{\rm FS} = 20$.
The temperature profile remains unchanged relative to model noFS, whereas the magnetic field strength $B_z$
is adjusted to ensure transverse force balance.
Figure~\ref{density_contour} displays how
	the thermal electron density
    is distributed in the $x-y$ plane (the left column)
    and the $y-z$ cut through $x=0$ (right)
    for both model noFS (the upper row) and model FS (lower).
When computing the GS emission, we assume that nonthermal electrons
    exist only inside the loop, namely in the cylinder with the nominal loop radius.     
We restrict ourselves to a line of sight (LoS) that is in the $x=0$ plane and makes
   an angle of $45^\circ$ with the $z$-axis.
Consequently, only the red-lined segments in Figure~\ref{density_contour} 
   are of interest, and they are $\sim 14$~Mm in length.       
However, we distinguish between two orientations where
   the observer on the Earth is placed at the opposite directions.
We refer to the two situations as ``LoS+'' and ``LoS-'', respectively.
The coordinate along a LoS, $z^{\prime}$,
   increases away from the observer. 
    
The system in both models is perturbed by a radially directed axisymmetric initial velocity
prescribed by
\begin{eqnarray}
v_x(x,y,z;t=0)=v_0\frac{r}{\sigma_r}{\rm exp}\left[\frac{1}{2}\left(1-\frac{r^2}{\sigma_r^2}\right)\right]
{\rm sin}\left(\frac{\pi z}{L}\right)\left(\frac{x}{r}\right)
\end{eqnarray} 
and 
\begin{eqnarray}
v_y(x,y,z;t=0)=v_0\frac{r}{\sigma_r}{\rm exp}\left[\frac{1}{2}\left(1-\frac{r^2}{\sigma_r^2}\right)\right]
{\rm sin}\left(\frac{\pi z}{L}\right)\left(\frac{y}{r}\right),
\end{eqnarray} 
where $v_0=10{\rm~km~s^{-1}}$ is the velocity amplitude,
and $\sigma_r=5$~Mm characterizes the spatial extent of the initial perturbation.
An axial fundamental sausage oscillation is established in both models,
   despite the transverse fine structuring in model FS.  
The fine strands nonetheless experience some kink-like motion
   (see \citetalias{2021ApJ...921L..17G} for details).
Figure~\ref{B_N_los} shows the 
    thermal electron number density ($N_e$) and the magnetic field strength ($B$) 
    as functions of $z^{\prime}$ and $t$ at orientation LoS+
    for both model noFS (the left column) and model FS (right).
The variations of $N_e$ and $B$ relative to the equilibrium values (at $t=0$)
    are also given. 
One sees for model noFS that the temporal variations of $B$ at
    all $z^{\prime}$ are in-phase, whereas those of $N_e$ outside the interval
    $2~{\rm Mm} \lesssim z^{\prime} \lesssim ~12~{\rm Mm}$ are in anti-phase with the variations in this interval. 
Moving on to model FS, one still readily discerns 
    periodic variations in both $N_e$ and $B$. 
In fact, their profiles for $z^{\prime} \gtrsim~8$~Mm are strikingly similar to
    what happens in model noFS.  
The most obvious differences between the two models lie in the range
    $z^{\prime} \lesssim ~8$~Mm where fine structures are present and move in a rather
    complicated manner.
These fine structures may move back and forth along the LoS in response
    to the dominant sausage oscillation.
Occasionally, they may deviate away from the LoS as a result of their
    kink-like behavior as well.    

\section{Gyrosynchrotron Emission: Reference Computations} %%%%%%%%%%%%%%%%%%%%%%%%%%%%%%%%%%%%%%%%
\label{S-results} 
We compute GS emission using the fast GS code
    (FGS) developed by \citet{2011ApJ...742...87K} (see also \citealt{2010ApJ...721.1127F}).
In short, FGS computes the local values of the absorption coefficient and emissivity,
    thereby accounting for inhomogeneous sources by integrating the radiative transfer equation.
Similar to \cite{2014ApJ...785...86R}, we assume that the number density of
    nonthermal electrons ($N_b$) is proportional to the thermal one ($N_e$),
    and specifically takes the form $N_b = 0.005N_e$.
Here the constant of proportionality is such that 
    the resulting $N_b$ is compatible with observations of typical flares \citep[e.g.,][]{2022Natur.606..674F}.
The spectral index of nonthermal electrons is $\delta=3.5$,
    with the energy range being $0.1$ to $10$~MeV.
The pitch angle distribution of nonthermal electrons is
    taken to be isotropic.
We assume that both LoS+ and LoS-
    thread a beam with a cross-sectional area of 48 km$\times$48 km
    when projected onto the plane of sky,
    in view of the spatial resolution that the Atacama Large Millimeter/submillimeter Array (ALMA) can achieve \citep{2016SSRv..200....1W}.
    
\begin{figure}    %%%%%%%%%%%%%%%%%% FIGURE 3
\begin{center}
	\includegraphics[width=0.8\textwidth,clip=]{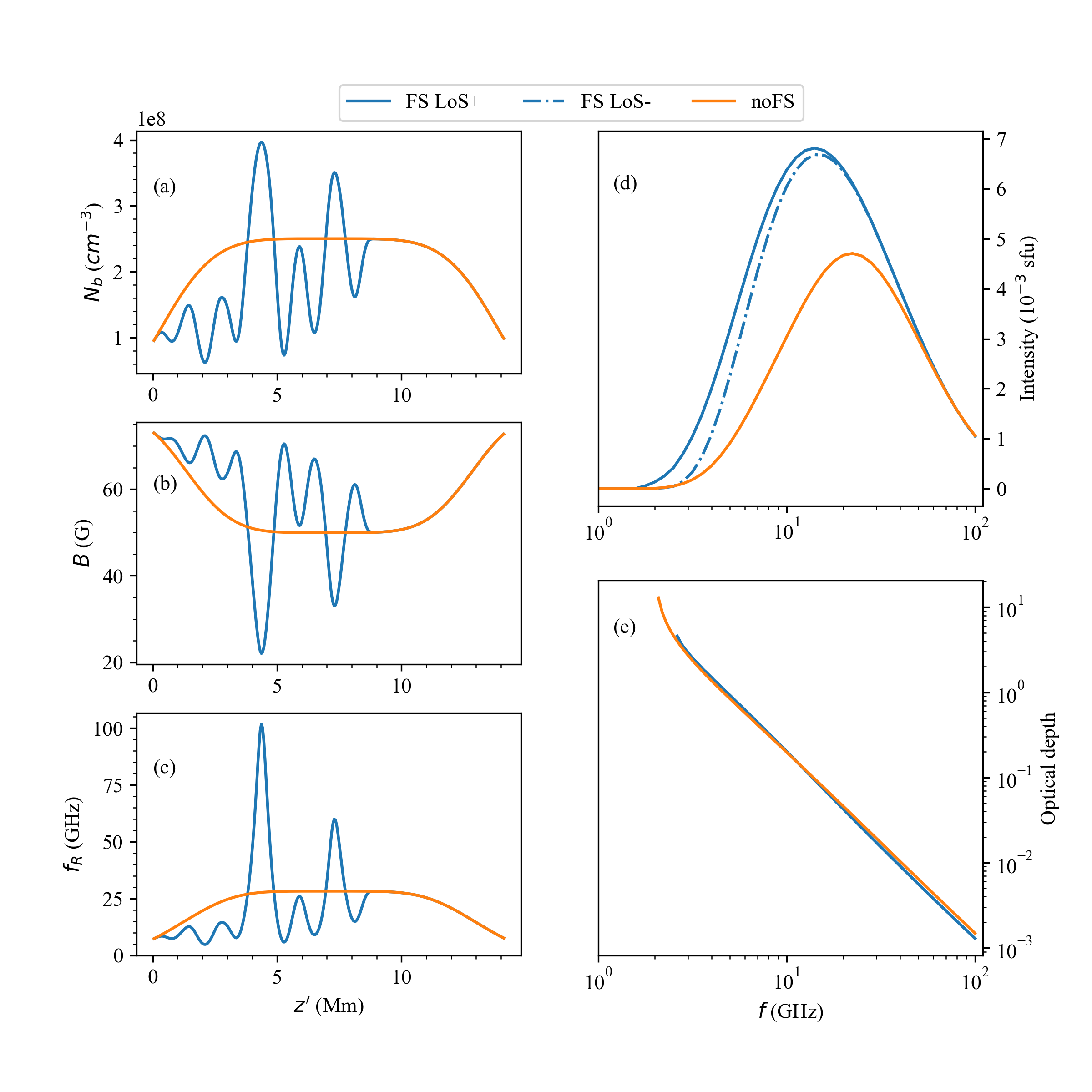}
	\caption{
	(a) The number density of nonthermal electrons ($N_b$),
	(b) the magnetic field strength ($B$),
	and (c) the Razin frequency ($f_R$) along the LoS. 
	(d) The intensity profiles. 
	(e) The optical depth averaged between the X and O modes.
	For model FS, the results for LoS+ and LoS- are discriminated by the 
	solid and dash-dotted lines in (d) and (e).}
	\label{emission_0}
	\end{center}
\end{figure}

Figure~\ref{emission_0} shows (a) the number density of nonthermal electrons $N_b$,
    (b) the magnetic field strength $B$,
    and (c) the Razin frequency $f_R$ 
    as a function of $z^{\prime}$ at $t=0$ for orientation LoS+
    for both model noFS (the orange curves) and model FS (blue).
The pertinent profiles for orientation LoS- can be readily deduced. 
Here $f_R$ is evaluated as $20N_e/|B_\perp|$ \citep[in CGS units, see][]{2014ApJ...785...86R}, with $B_\perp$
    being the instantaneous $B$ component transverse to the LoS.    
The relevance of $f_R$ is that, when the thermal electron density is high, 
    the spectral peak of GS emission is formed due to the Razin effect,
    which considerably suppresses the intensity at frequencies below $f_R$ \citep{razin1960theory}.
Figure \ref{emission_0}d shows the spectral profiles for model
	noFS (the orange curve) and model FS (blue) at $t=0$.
We discriminate the profiles for LoS+ and LoS- by the solid and dash-dotted curves
	only for model FS.
For model noFS, the results are identical at both orientations 
	due to the symmetry of this LoS.
Both the spectral profile and the peak frequency of model FS
	are different from the results of model noFS.
These differences are caused by the physical parameters 
	(e.g., Figures~\ref{emission_0}a, \ref{emission_0}b) and Razin frequencies (Figure~\ref{emission_0}c) along the LoS.
Observing the loop at LoS+ or LoS- for model FS, 
	one sees that the intensity does not change at high frequencies,
	but changes at low frequencies.
Figure \ref{emission_0}e shows the optical depth $\tau$
	as a function of frequency $f$ at $t=0$ for both models.
Here $\tau$ is obtained by integrating the local $\varkappa$
	along the LoS,
	with $\varkappa$ being the average of the absorption coefficients
	of the X and O modes weighted by their local intensities.
Similar to Figure \ref{emission_0}(d), we discriminate $\tau$ of LoS+ and LoS- only for model FS,
	though we do not see any substantial difference for the two orientations.
From Figure \ref{emission_0}(e), one sees that at frequencies $f\gtrsim5$~GHz, 
	the optical depth is less than unity and thus the loop is optically thin for both models.

\begin{figure}    %%%%%%%%%%%%%%%%%% FIGURE 2
	\begin{center}
		\includegraphics[width=\textwidth,clip=]{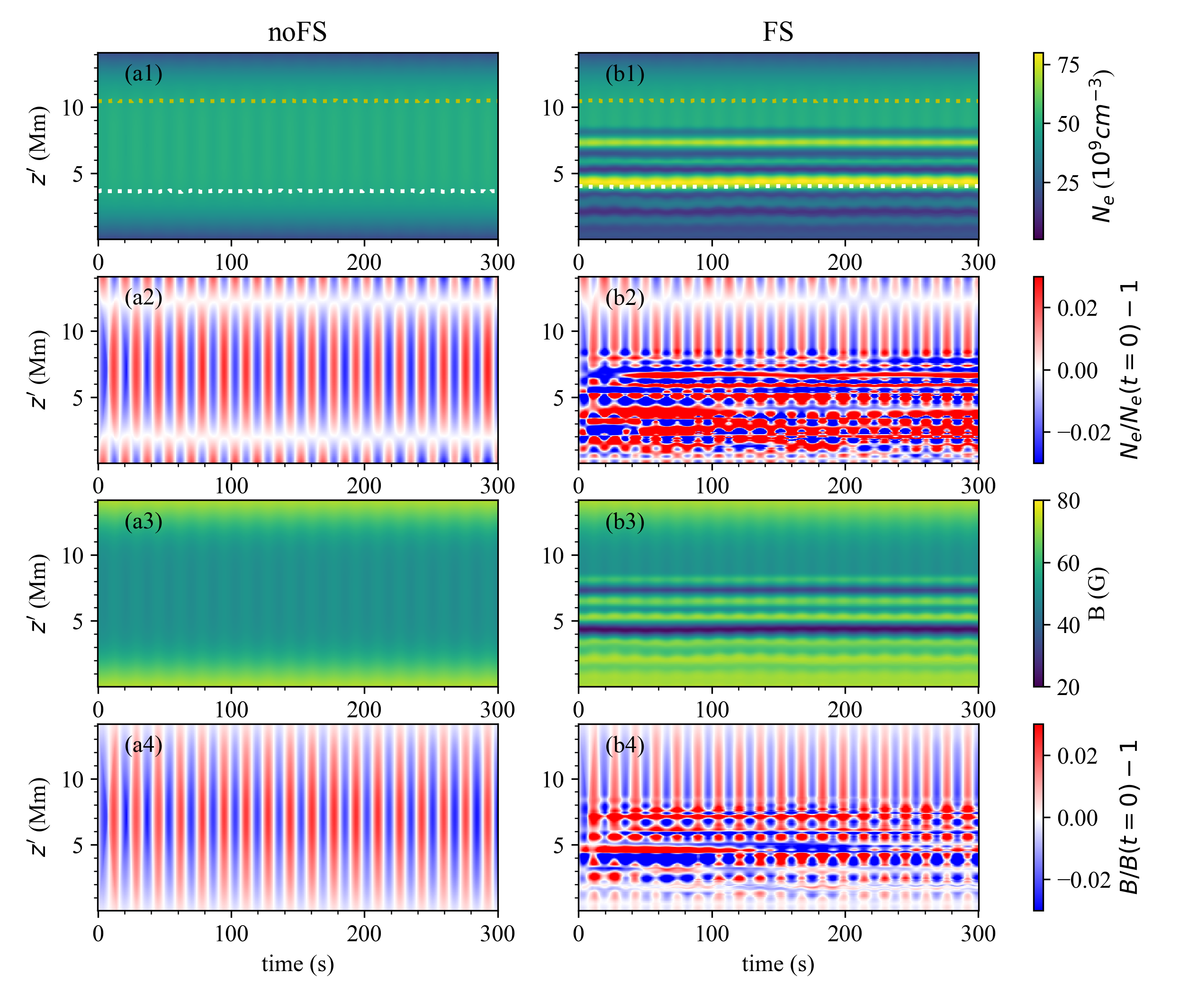}
		\caption{Left: Temporal variations of (a1) the electron number density ($N_e$), (a2) the relative variation of $N_e$,
			  (a3) the magnetic field strength ($B$), and (a4) the relative variation of $B$ along the LoS for model noFS.
			  Right: Same as left but for model FS.
			  The dotted curves in (a1) and (b1) mark the locations where the optical depth of the 2.5~GHz emission reaches unity
			  for LoS+ (white) and LoS- (yellow).}
		\label{B_N_los}
	\end{center}
\end{figure}

\begin{figure}    %%%%%%%%%%%%%%%%%% FIGURE 4
	\begin{center}
		\includegraphics[width=\textwidth,clip=]{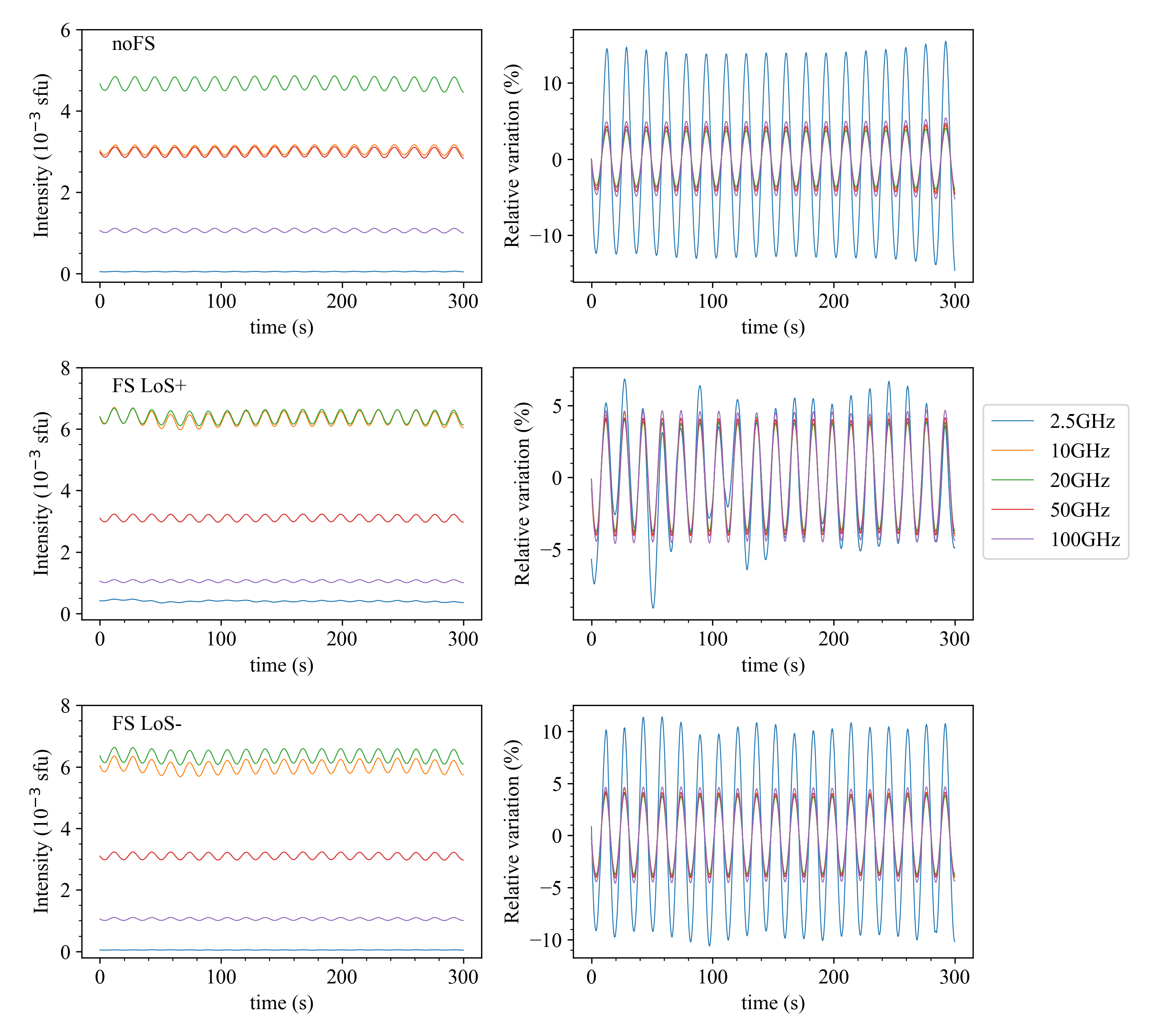}
		
		\caption{Intensity (left) and its relative variation (right) at five frequencies
			for model noFS (top), model FS (LoS+, middle), and model FS (LoS-, bottom).
			The curves of relative variation are smoothed with a window of 2 periods.}
		\label{int_evolution}
	\end{center}
\end{figure}

Figure~\ref{int_evolution} shows the temporal evolution 
	of intensity (left column) and the variations with respect to $t=0$ (right),
	taking the results at five frequencies as examples.
For model noFS (top row), the intensities at all frequencies, 		
	including the optically thick 2.5~GHz, 
	oscillate in phase at the period of sausage mode.
The result that low and high frequency emissions oscillate in-phase
	when the Razin effect dominates agrees with that of \cite{2007ARep...51..588R} and \cite{2012ApJ...748..140M}
	where a homogeneous emission source was assumed.
This result also agrees with \cite{2014ApJ...785...86R} 
	at some viewing angles when an inhomogeneous emission source was examined.
At high frequencies, the loop is optically thin,
	thus every voxel along the LoS contributes to the total emission,
	making the intensity variation follow the oscillation of the global sausage mode.
At low frequencies (e.g., 2.5~GHz), however, the loop is optically thick,
	so the emission mostly comes from the volume
	where the optical depth is about unity (see the dotted lines in Figure \ref{B_N_los}).
It is inappropriate to compare our results with the
	approximate formula proposed by \cite{1982ApJ...259..350D}, 
	where the Razin effect was not considered.
Furthermore, the physical parameters vary smoothly at the loop boundary,
	making the intensity of the optically thick emission more difficult to estimate
	using these approximate formula.
For model FS (middle and bottom rows), the intensities at all frequencies 
	also oscillate in-phase.
The most striking difference, relative to model noFS,
	is the relative intensity variation of $2.5$~GHz.
The modulation amplitude at this frequency is reduced by the fine structures,
	with the reduction being more significant for LoS+.
This effect is attributed to the optical thickness.
In the optically thick regime, the intensity is dominated by the layer where 
	the optical depth reaches unity,
	thus the fine structures or the orientation observing them would influence
	the intensity and its variation.
At high frequencies, the intensity variations are not 
	obviously influenced by the fine structures or the viewing angle.
%Our results indicate that the fine structures can influence the relative variation of intensity at the optical thick low frequency,
%but at high frequency, this effect is marginal.

\begin{figure}    %%%%%%%%%%%%%%%%%% FIGURE 5
	\begin{center}
		\includegraphics[width=0.6\textwidth,clip=]{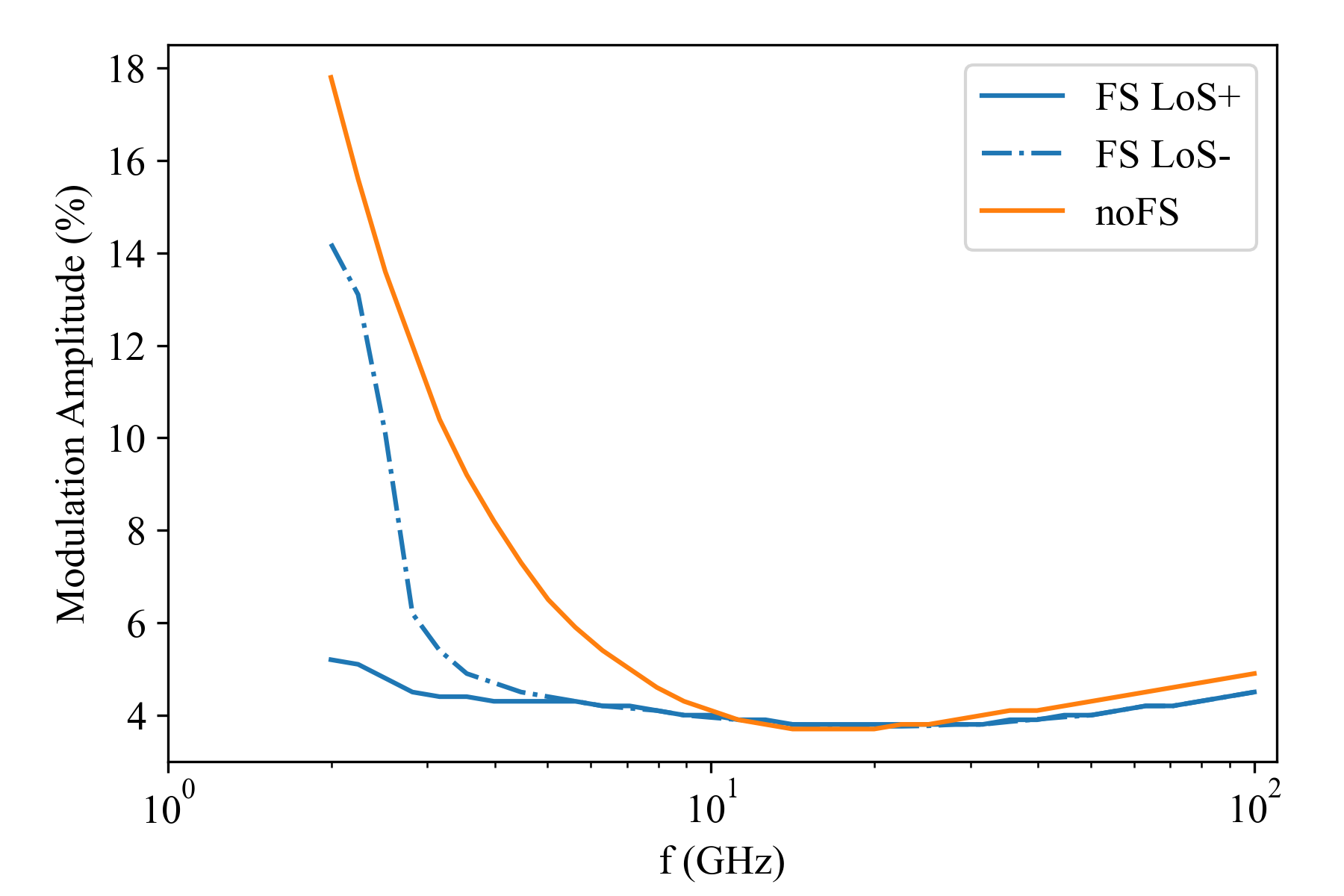}
		\caption{Modulation amplitudes at different frequencies for models FS and noFS.}
		\label{modulation_amplitude}
	\end{center}
\end{figure}

Figure~\ref{modulation_amplitude} plots the modulation amplitude 
	as a function of frequency for both models. 
We obtain the modulation amplitude at each frequency by fitting the 
	relative variation (the right column in Figure~\ref{int_evolution}) 
	with a sinusoidal curve.
For model noFS, the modulation amplitude reaches its minimum 
	around the peak frequency, and increases rapidly with decreasing frequency,
	consistent with the results of \cite{2012ApJ...748..140M} and \cite{2015SoPh..290.1173K}.
For model FS, the modulation amplitudes at low frequencies are
	dramatically reduced, with the effect being more pronounced for LoS+. 
At high frequencies, the modulation amplitudes are not 
	obviously influenced by the fine structures.
This effect is also due to different optical depths 
	at different frequencies.
As an example, we plot in Figure~\ref{B_N_los} the positions 
	where the optical depth of the 2.5~GHz emission 
	reaches unity for LoS+ (the white curves) and LoS- (yellow).
For LoS+, the emission is dominated by the finely structured region,
	which destructs the coherent variations of physical parameters along the LoS
	and hence leads to the decrease of the modulation amplitude.
For LoS-, however, the emission coming from the structuring 
	is greatly reduced and the contribution from the region of coherent variations dominates.
The end result is that
	the modulation amplitudes are larger at LoS- than at LoS+
	for low-frequency emissions where the loop is optically thick. 
	
\section{Gyrosynchrotron Emission: Further computations}
\label{further computations}
The resultant GS emission depends on multiple parameters that characterize, say, the magnetic field strength,
	the nonthermal electrons, and the background thermal plasma.
The ideal way to examine the influences of these parameters is to perform
	a parametric study with only one parameter changed each time.
However, the inclusion of density fine structures also
	introduce structuring on such parameters as the magnetic field strength in model FS,
	due to the force-balance condition.
In addition, the randomly distributed fine structures does not guarantee that the 
	total number of the nonthermal electrons remains the same as in model noFS.
In this section, 
 	we thus perform further MHD simulations and forward modeling analyses
 	besides model noFS and model FS.
Table \ref{table1} summarizes the details of all models.

\begin{table}[h!]
	\begin{center}
		\caption{Model description}
		\label{table1}
		\begin{threeparttable}
		\begin{tabular}{l c c}
			\hline
			\hline
			\textbf{Model} & \textbf{Fine-structured?} & {\textbf{Nonthermal electrons}} \\
			\hline
			noFS      & No            & \multirow{3}*{\makecell{$N_b = 0.005N_e$, $E=[0.1,10]$~MeV \\ SPL\tnote{1}, $\delta=3.5$}}\\
			\cline{1-2}		
			FS        & $\rho$, $B_z$ & ~\\
			\cline{1-2}
			FS\_conB  & $\rho$, $T$   & ~\\
			\cline{1-3}
			FS\_conNb & $\rho$, $B_z$ & \makecell{$N_b = 0.0055N_e$, $E=[0.1,10]$~MeV \\SPL, $\delta=3.5$}\\
			\cline{1-3}
			FS\_DPL   & $\rho$, $B_z$ & \multirow{2}*{\makecell{$N_b = 0.005N_e$, $E=[0.01,10]$~MeV \\DPL\tnote{2}, $\delta_1=1.5$, $\delta_2=3.5$}}\\
			\cline{1-2}
			noFS\_DPL & No              & ~\\
			\hline
		\end{tabular}
	    \begin{tablenotes} 
	    	\footnotesize              
		    \item[1] Single power law \item[2] Double power law
     	\end{tablenotes}       
		\end{threeparttable}
	\end{center}
\end{table}
	
\subsection{Magnetic flux}
\begin{figure}    %%%%%%%%%%%%%%%%%% FIGURE 6
	\begin{center}
		\includegraphics[width=\textwidth,clip=]{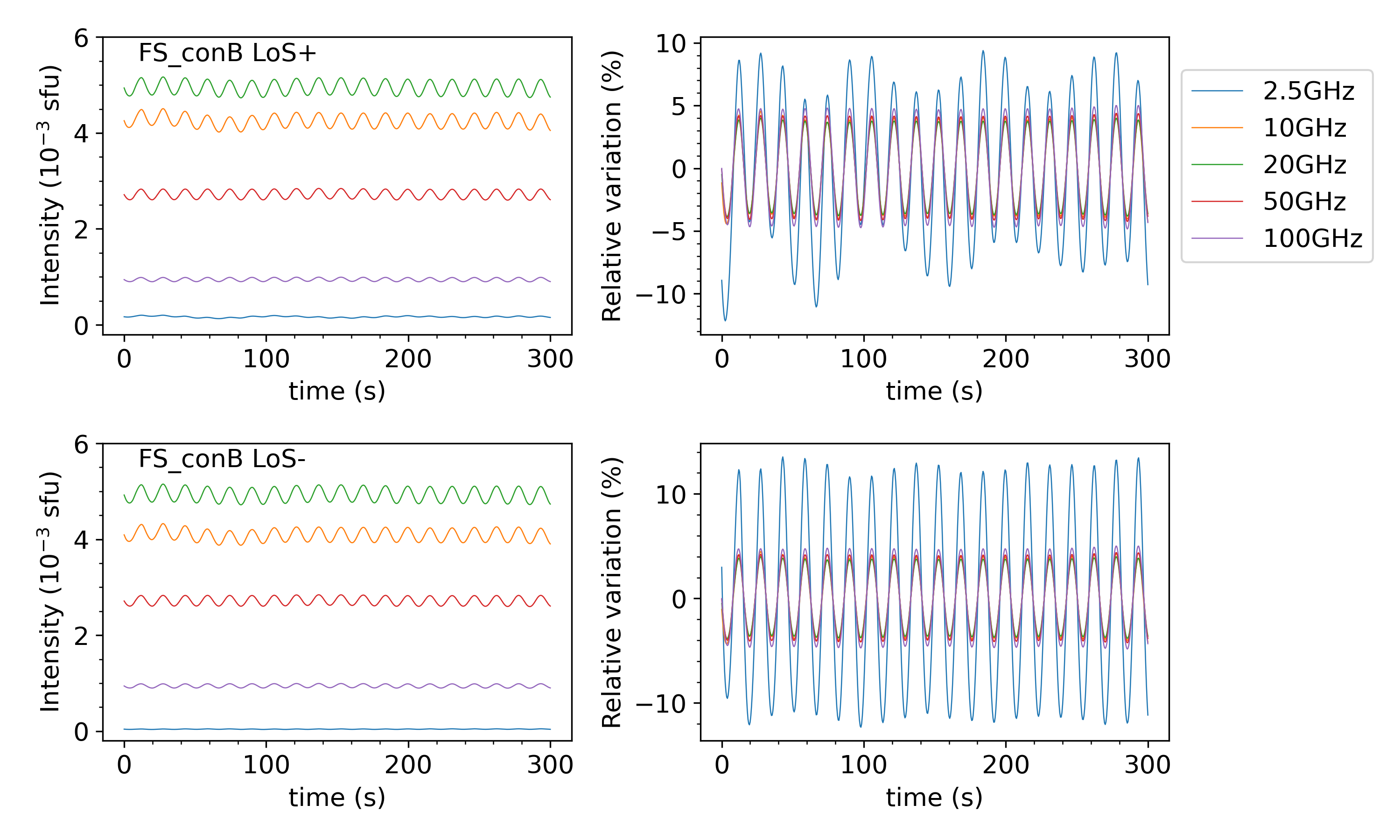}
		\caption{Similar to Figure \ref{int_evolution} but for model FS\_conB.}
		\label{int_evolution_conB}
	\end{center}
\end{figure}

In model FS, the magnetic field strength is also fine-structured and the total magnetic flux 
	is not the same as in model noFS.
We conduct another MHD simulation, to be labeled `FS\_conB',
	where the equilibrium magnetic field has the same profile as in model noFS.
Force balance in model FS\_conB is maintained in a way that
	density structuring is counteracted by the temperature structuring.
Figure \ref{int_evolution_conB} shows the emission intensities and their relative variations
	at five frequencies.
Though the intensities are slightly different when compared with model FS,
	the relative variations are quite similar. 
Figure \ref{modulation_amplitude_conB} compares the modulation amplitudes
	of model FS\_conB and model FS.
We find that the modulation amplitudes are different at some frequencies,
	whereas the general trend remains the same as model FS.
	
\subsection{Number of the nonthermal electrons}
\begin{figure}    %%%%%%%%%%%%%%%%%% FIGURE 8
	\begin{center}
		\includegraphics[width=\textwidth,clip=]{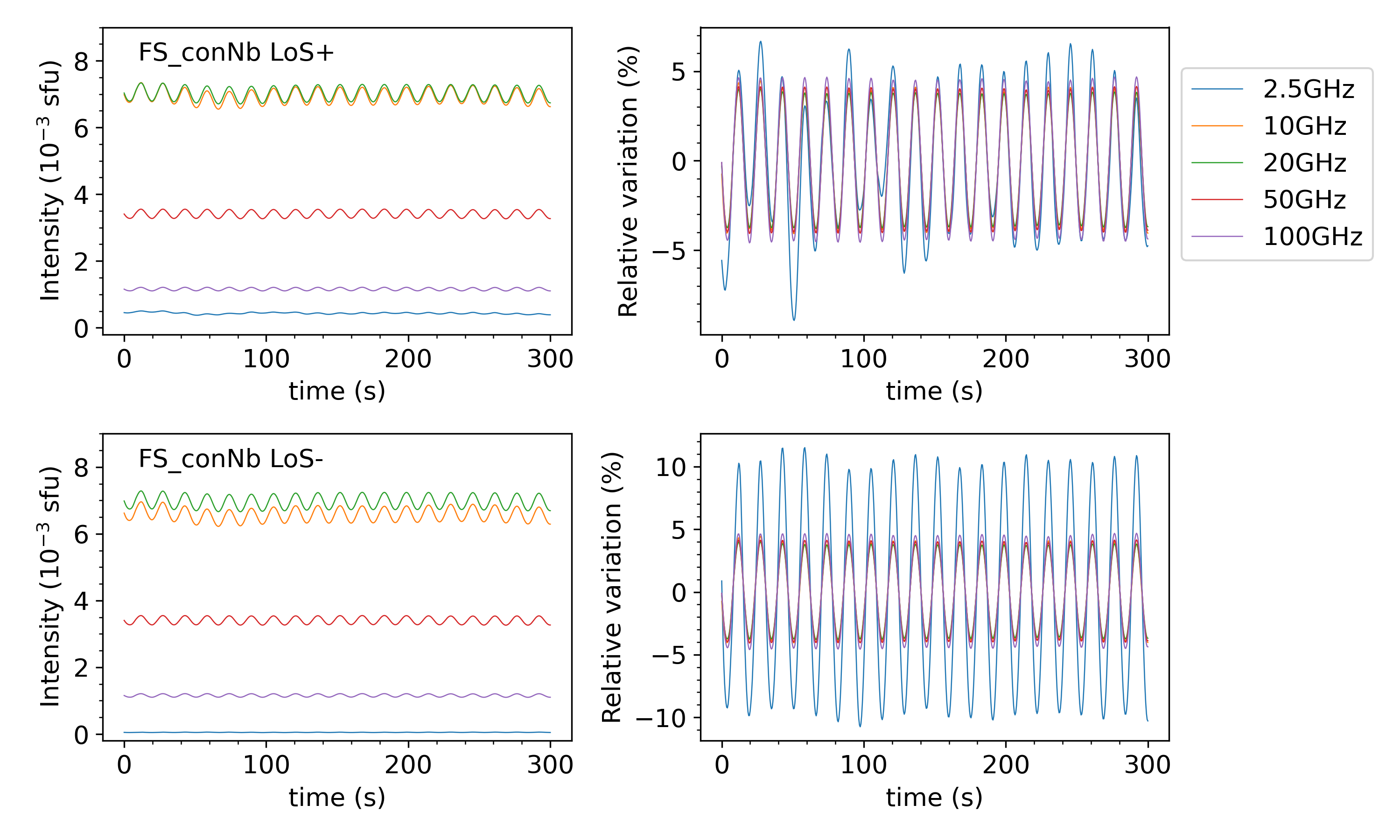}
		\caption{Similar to Figure \ref{int_evolution} but for model FS\_conNb.}
		\label{int_evolution_conNb}
	\end{center}
\end{figure}

When computing the GS emission, we assume that the number of the nonthermal electron 
	is proportional to that of the thermal one, i.e., $N_b = 0.005N_e$.
The resultant number of the nonthermal electron inside the loop is 8.5\% lower in model FS than in model noFS.
To make sure that the total number of the nonthermal electron in model FS is exactly the same as in model noFS, 
	we perform another forward modeling computation, to be labeled as `FS\_conNb',
	where $N_b = 0.0055N_e$ is assumed.
Figures \ref{int_evolution_conNb} and \ref{modulation_amplitude_conNb} present the results of model FS\_conNb.
One sees that the intensity variations and the modulation amplitudes are quite similar 
	to model FS. 

\subsection{Lower energy cut-off for nonthermal electrons}
\begin{figure}    %%%%%%%%%%%%%%%%%% FIGURE 10
	\begin{center}
		\includegraphics[width=\textwidth,clip=]{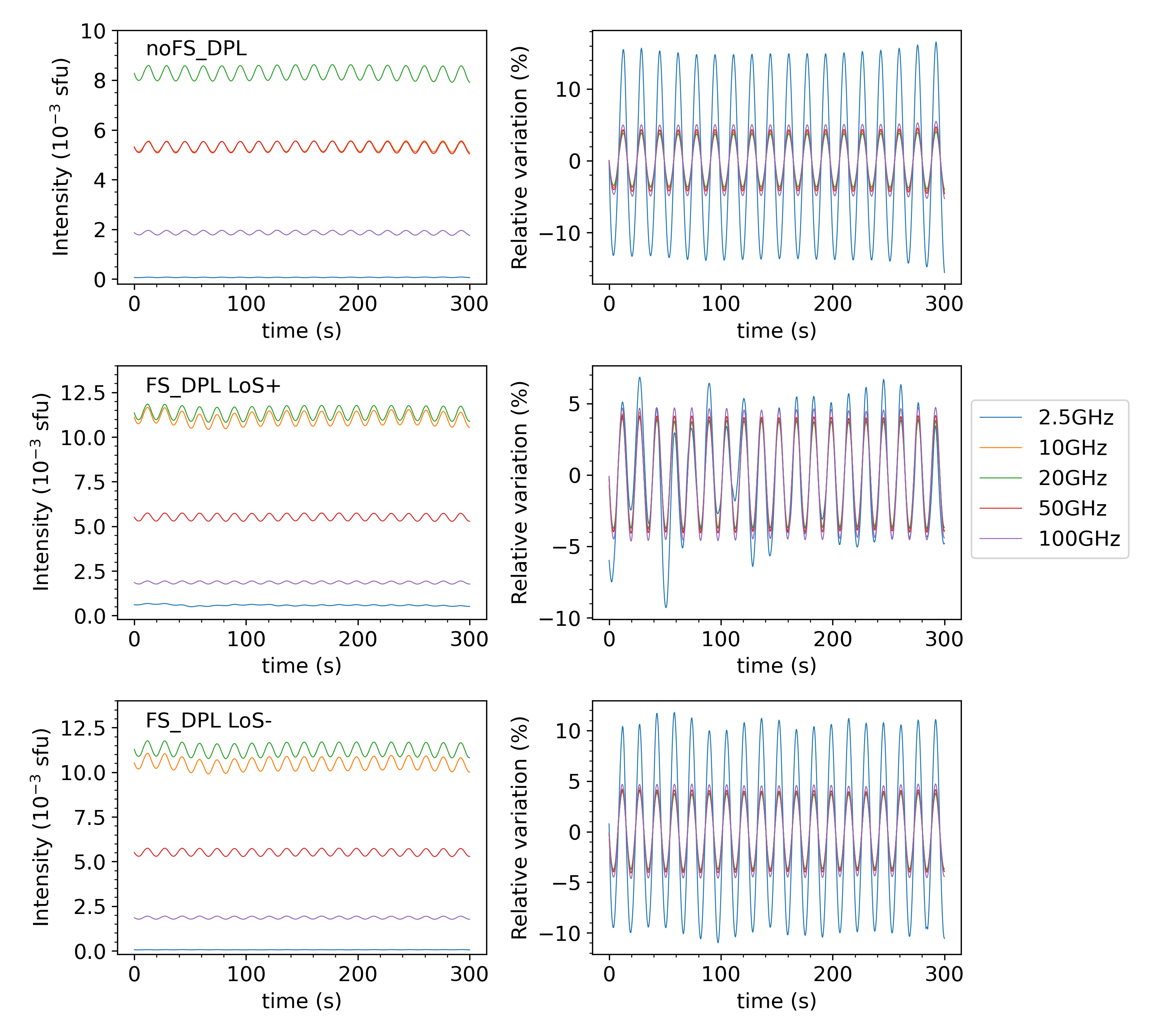}
		\caption{Similar to Figure \ref{int_evolution} but for model noFS\_DPL and FS\_DPL.}
		\label{int_evolution_DPL}
	\end{center}
\end{figure}

\begin{figure}    %%%%%%%%%%%%%%%%%% FIGURE 11
	\begin{center}
		\includegraphics[width=0.6\textwidth,clip=]{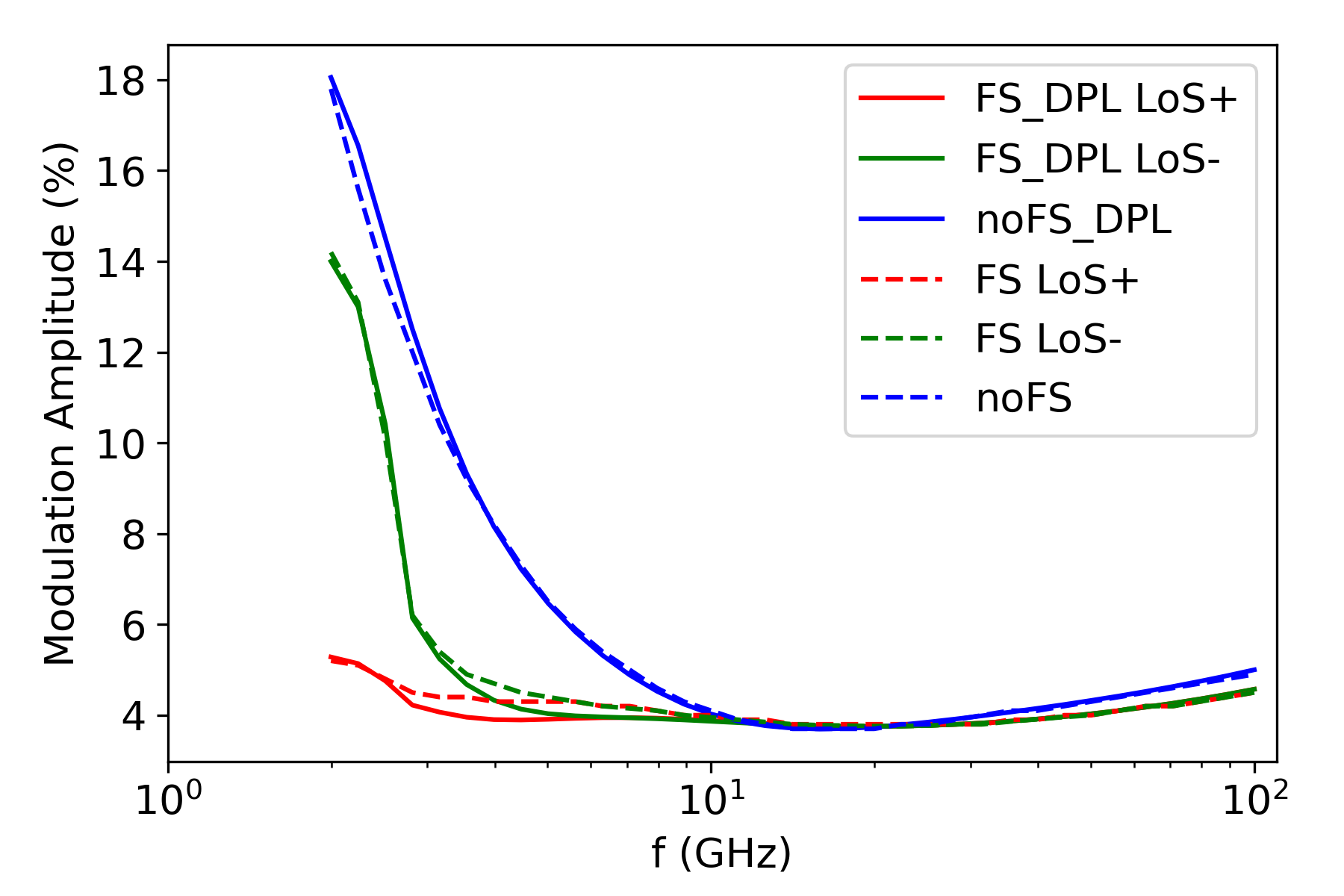}
		\caption{Modulation amplitudes for models with single power law (the dashed lines) and double power law (solid).}
		\label{modulation_amplitude_DPL}
	\end{center}
\end{figure}

In the above mentioned forward modeling analysis,
	the lower energy cut-off of the nonthermal electrons is $E_{\rm min}=0.1$~MeV,
	which is higher than the typical value inferred from hard X-ray observations.
The electrons with lower energies can influence the self-absorption at low frequencies,
	but their contribution to the GS emission is usually negligible \citep{2003ApJ...586..606H}.
To address the influence of the lower energy nonthermal electrons in our model,
	we perform additional forward modeling analysis,
	to be labeled as noFS\_DPL and FS\_DPL.
For these two models, the energy of the nonthermal electrons ranges from $E_{\rm min}=0.01$~MeV to $E_{\rm max}=10$~MeV.
The spectrum of the nonthermal electrons takes a double power law shape with a break at $0.5$~MeV,
	with the spectral index being $\delta_1=1.5$ in the low energy band and $\delta_2=3.5$ in the high energy band,
	similar to \cite{2014ApJ...785...86R}.
Figure~\ref{int_evolution_DPL} shows the intensities and their variations.
Relative to Figure~\ref{int_evolution},
 one finds that the intensities are somehow different while the relative variations are similar.
Figure~\ref{modulation_amplitude_DPL} compares the modulation amplitudes for models FS and noFS.
We find that the influence of the lower energy nonthermal electrons
 	is minor on the modulation amplitudes of the GS emission in our model.

\section{Summary and Conclusion}
\label{S-summary}
\begin{figure}    %%%%%%%%%%%%%%%%%% FIGURE 1
\begin{center}
	\includegraphics[width=\textwidth,clip=]{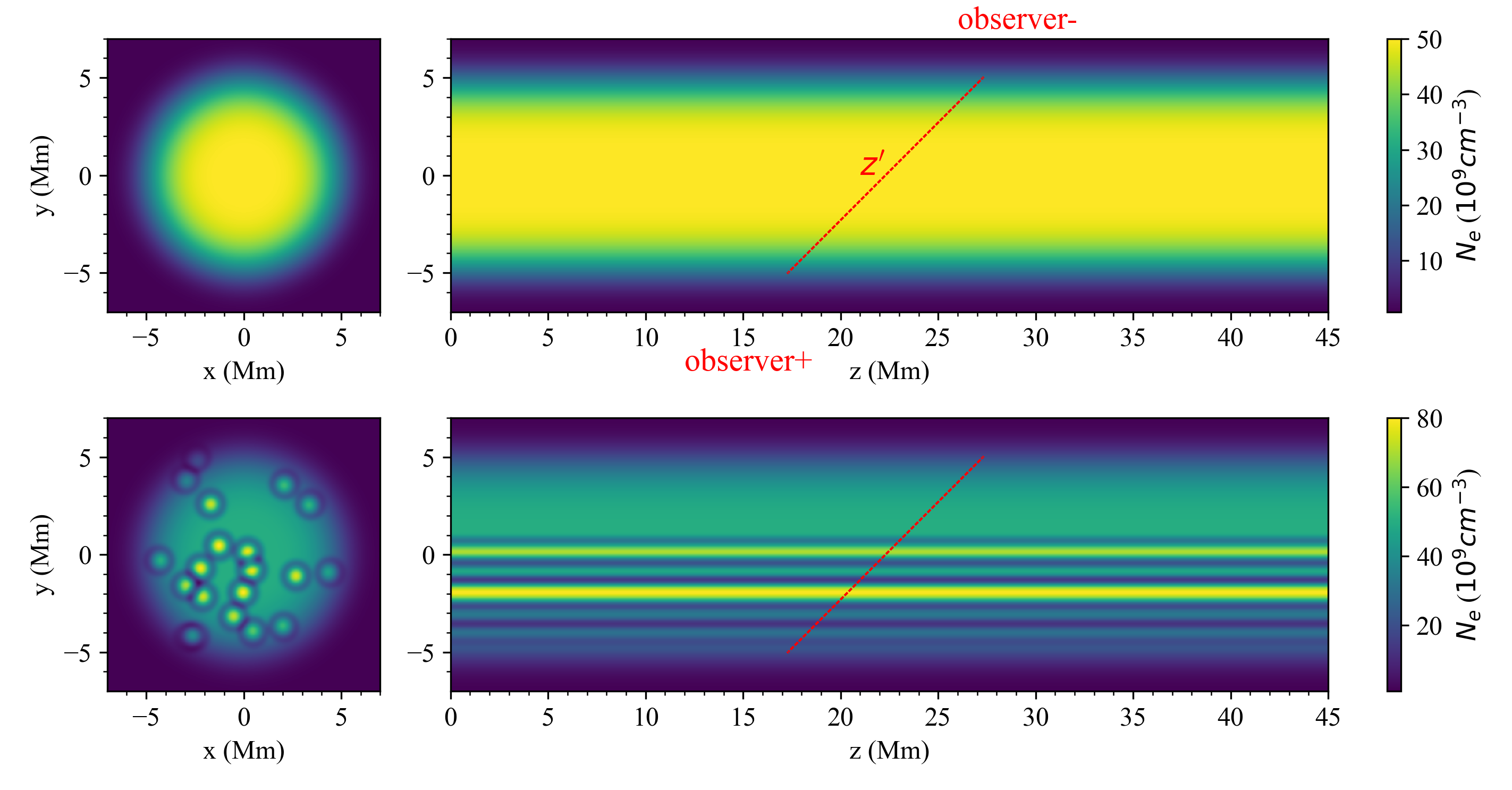}
	\caption{Distribution of the thermal electron number density ($N_e$) at $t=0$ for model noFS (top) and model FS (bottom) 
		in the $x-y$ (left, with $z=0$) and $y-z$ (right, with $x=0$) planes. 
		The red dotted lines mark the line of sight (LoS) projected onto the $y-z$ plane. Two LoS orientations are indicated by ``observer+'' (LoS+) and ``observer-'' (LoS-), respectively.}	
	\label{density_contour}
\end{center}
\end{figure}

Using three-dimensional MHD models,
	we study the influence of fine structures on the GS emission
	of flare loops modulated by axial fundamental sausage modes.
The numerical models used for our reference computations
	are the same as those in \citetalias{2021ApJ...921L..17G},
	where two MHD models, namely noFS and FS, are simulated.
Model noFS sees an equilibrium flare loop as a density-enhanced
	monolithic cylinder,
	whereas model FS modifies noFS by randomly introducing transverse fine structures to the loop interior.	
In both models, sausage modes are excited by an axisymmetric 
	velocity perturbation.
We compute GS emissions using the fast GS codes
	\citep{2010ApJ...721.1127F,2011ApJ...742...87K}.
We assume that the number density of the nonthermal electrons 
	is proportional to that of the thermal electrons.
The spectral index of the nonthermal electrons is 3.5,
	with its energy range being 0.1 MeV to 10 MeV.
The pitch angle distribution is assumed to be isotropic.
The temporal variation of the emission intensity is analyzed for 
	a LoS crossing the loop center (red lines in Figure \ref{density_contour}).
We find that the low and high frequency intensities oscillate in-phase
	at the period of sausage mode for both model noFS and model FS.
Fine structures only influence the intensity variation of the 
	low-frequency emissions,
	dramatically reducing the
	modulation amplitudes.
How significant this effect may be also depends on the orientation of the observer (i.e., LoS+ or LoS-).
At high frequencies, the modulation amplitudes are not
	obviously influenced by the fine structures.
Further computations (see Table \ref{table1}) with different MHD models or nonthermal electron distributions
	are also examined,
	the results being largely similar to the reference computations.
These results are helpful for understanding the GS emission modulated 
	by sausage modes in flare loops, as well as for identifying sausage modes using radio observations.
Combining our forward modeling results with the MHD simulations of
	\citetalias{2021ApJ...921L..17G},
	we conclude that the periodic oscillations of sausage modes 
	are not wiped out by the loop fine structures,
	and sausage modes are a promising 
	mechanism for interpreting flare QPPs even when flare loops are fine-structured.
		
\begin{figure}    %%%%%%%%%%%%%%%%%% FIGURE 7
	\begin{center}
		\includegraphics[width=0.6\textwidth,clip=]{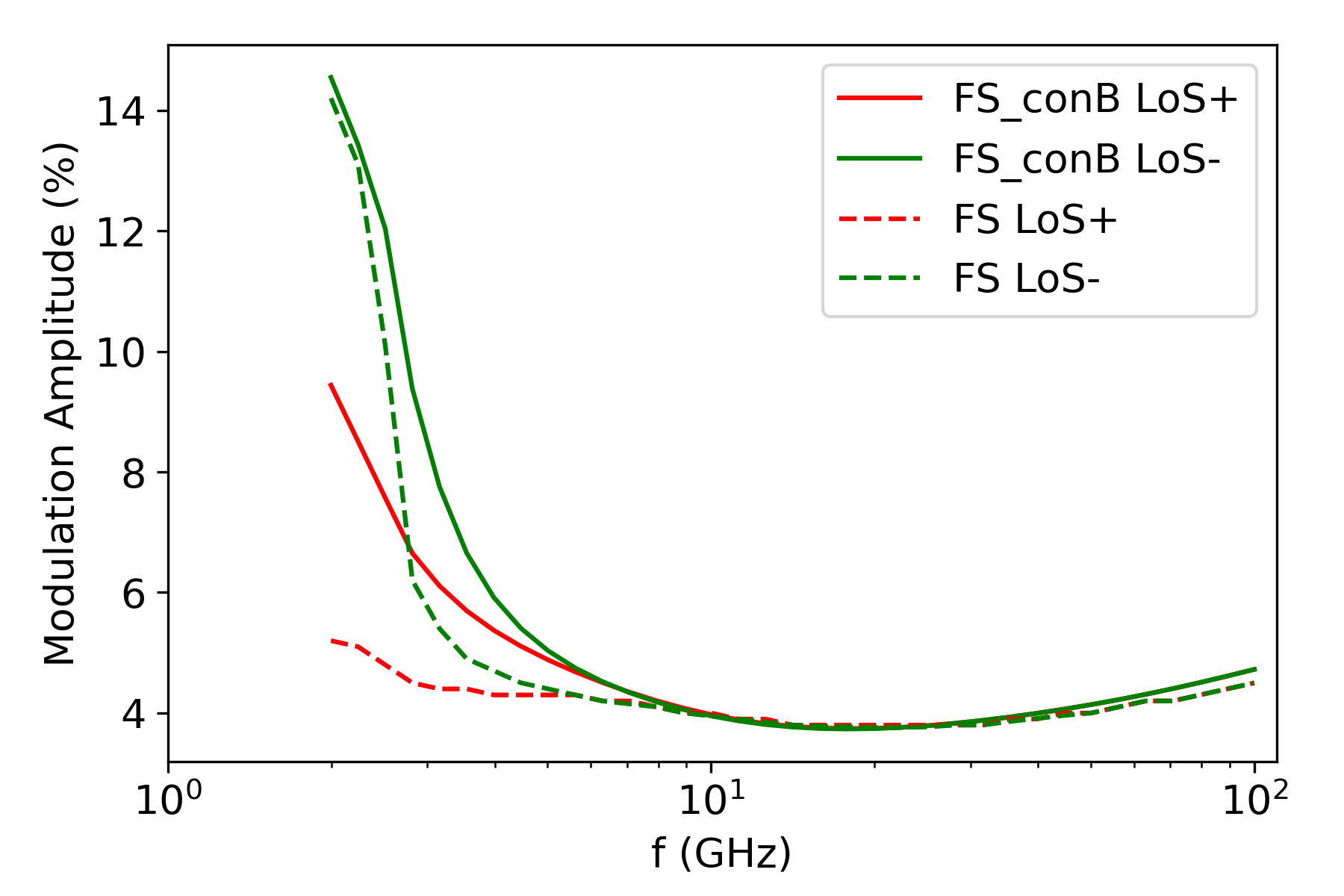}
		\caption{Modulation amplitudes of model FS (the dashed lines) and model FS\_conB (solid).}
		\label{modulation_amplitude_conB}
	\end{center}
\end{figure}

\begin{figure}    %%%%%%%%%%%%%%%%%% FIGURE 9
	\begin{center}
		\includegraphics[width=0.6\textwidth,clip=]{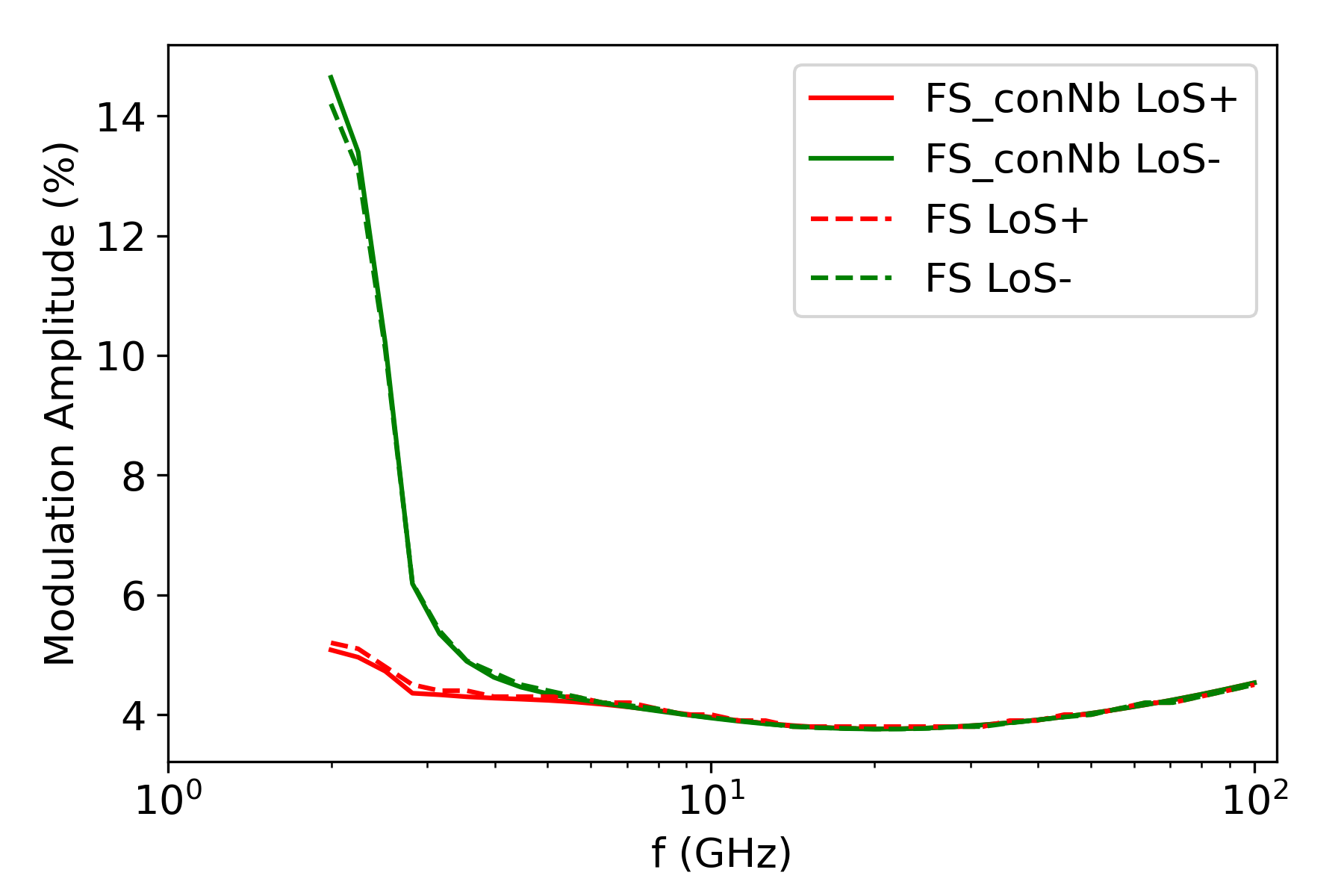}
		\caption{Modulation amplitudes of model FS (the dashed lines) and model FS\_conNb (solid).}
		\label{modulation_amplitude_conNb}
	\end{center}
\end{figure}

Our results show that the modulation amplitudes of the GS emission intensities at low frequencies
	are dramatically reduced by the loop fine structures.
	This effect is mainly attributed to the optical thickness of the low-frequency emissions.
	However, such parameters as the total number of non-thermal electrons or the total magnetic flux 
	can also influence the GS emissions and potentially affect the modulation amplitudes.
	The difference of the non-thermal electron numbers is 8.5\% between model FS and model noFS.
	The influence of this difference on the modulation amplitudes is marginal,
	as shown in Figure~\ref{modulation_amplitude_conNb}.
	The influence of the total magnetic flux is slightly stronger.
	From Figure~\ref{modulation_amplitude_conB} one sees some difference
	between model FS and model FS\_conB at low frequencies.
	Despite this difference,
	the modulation amplitudes of model FS\_conB are still obviously smaller than those of model noFS.
	These results demonstrate that the optical thickness is the dominant factor for reducing the modulation amplitudes
	of the low-frequency emission intensities.

Even though the modulation amplitudes at low frequencies 
	are significantly larger,
	the emission intensities at these frequencies are quite low,
	making their detections challenging.
This is because the emission at low frequencies is significantly
	suppressed by the Razin effect.
For flare loops with lower density, where the Razin suppression is not important,
	the intensity at low frequencies could be larger.
In this case, the peak frequency is determined by self-absorption,
	so all frequencies below the peak should be optically thick.
Our above mentioned effects relating to the optically thick regime
	are expected to be more pronounced and occur in a wider frequency range.

\begin{acknowledgments}
We thank the reviewer for his/her comments that help improve this manuscript.
This work is supported by the National Natural Science Foundation of China (41904150, 41974200, 11761141002).
We gratefully acknowledge ISSI-BJ for supporting the international team “Magnetohydrodynamic wavetrains as a tool for probing the solar corona ”.
\end{acknowledgments}
\clearpage

\clearpage

\clearpage

\clearpage

\clearpage

\clearpage

\clearpage

\clearpage

\end{document}